\documentclass [prl,reprint,superscriptaddress,showpacs]{revtex4-1}
\usepackage{graphicx}

\begin{document}

\title{Ultrafast modulation of the chemical potential in BaFe$_2$As$_2$ by coherent phonons}

\author{L.~X. Yang}
\author{G. Rohde}
\author{T. Rohwer}
\author{A. Stange}
\author{K. Hanff}
\author{C. Sohrt}
\affiliation{Institut f\"ur Experimentelle und Angewandte Physik, Christian-Albrechts-Universit\"at zu Kiel, D--24098 Kiel, Germany}
\author{L. Rettig}
\affiliation{Fakult\"at f\"ur Physik, Universit\"at Duisburg-Essen, D--47048 Duisburg, Germany}
\author{R. Cort\'{e}s}
\affiliation{Fachbereich Physik, Freie Universit\"at Berlin, D--14195, Germany}
\author{F.~Chen}
\author{D.~L.~Feng} 
\affiliation{Surface Physics Laboratory, Physics Department and Advanced Materials Laboratory, Fudan University, Shanghai 200433, P.\ R.\ China}
\author{T. Wolf}
\affiliation{Institut f\"{u}r Festk\"orperphysik, Karlsruhe Institute of Technology, D--76021 Karlsruhe, Germany}
\author{B. Kamble}
\author{I. Eremin}
\affiliation{Institut f\"ur Theoretische Physik III, Ruhr-Universit\"at Bochum, D--44801 Bochum, Germany}
\author{T. Popmintchev}
\author{M.~M.~Murnane}
\author{H.~C.~Kapteyn}
\affiliation{JILA, University of Colorado and NIST, Boulder, CO 80309-0440 Boulder, USA}
\author{L. Kipp}
\affiliation{Institut f\"ur Experimentelle und Angewandte Physik, Christian-Albrechts-Universit\"at zu Kiel, D--24098 Kiel, Germany}
\author{J. Fink}
\affiliation{Leibniz-Institut f\"ur Festk\"orper- und Materialforschung Dresden, D--01171 Dresden, Germany}
\author{M. Bauer}
\affiliation{Institut f\"ur Experimentelle und Angewandte Physik, Christian-Albrechts-Universit\"at zu Kiel, D--24098 Kiel, Germany}
\author{U. Bovensiepen}
\affiliation{Fakult\"at f\"ur Physik, Universit\"at Duisburg-Essen, D--47048 Duisburg, Germany}
\author{K. Rossnagel}
\affiliation{Institut f\"ur Experimentelle und Angewandte Physik, Christian-Albrechts-Universit\"at zu Kiel, D--24098 Kiel, Germany}


\begin{abstract}
Time- and angle-resolved extreme ultraviolet photoemission spectroscopy is used to study the electronic structure dynamics in BaFe$_2$As$_2$ around the high-symmetry points $\Gamma$ and $M$. A global oscillation of the Fermi level at the frequency of the $A_{1g}$(As) phonon mode is observed. It is argued that this behavior reflects a modulation of the effective chemical potential in the photoexcited surface region that arises from the high sensitivity of the band structure near the Fermi level to the $A_{1g}$ phonon mode combined with a low electron diffusivity perpendicular to the layers. The results establish a novel way to tune the electronic properties of iron pnictides: coherent control of the effective chemical potential. The results further suggest that the equilibration time for the effective chemical potential needs to be considered in the ultrafast electronic structure dynamics of materials with weak interlayer coupling.
\end{abstract}


\maketitle

Time-resolved optical and photoemission spectroscopies have become important tools to probe the microscopic details of electron-phonon coupling. The prime example is the reliable determination of the coupling parameter from measured relaxation times of excited electrons \cite{Allen_87,Brorson_90,Gadermaier_10,Stojchevska_10,Rettig_13,Johannsen_13}. More recently, time-resolved spectroscopies have provided novel insights into the transient behavior of electronically ordered phases, specifically charge-density waves and superconductivity, in which electron-phonon coupling plays a prominent role \cite{Perfetti_06,Schmitt_08,Kusar_08,Tomeljak_09,Cortes_11,Petersen_11,Hellmann_12,Smallwood_12}.

A particularly intriguing aspect of electron-phonon coupling often observed in pump-probe spectroscopy is the generation of coherent optical phonons \cite{Ishioka_10} and their subsequent modulation of electronic properties. This effect not only provides a powerful means to study femtosecond lattice dynamics \cite{Bothschafter_13}, but can also be used to coherently control the electronic structure of materials. Through time- and angle-resolved photoemission spectroscopy (trARPES), coherent phonon-induced oscillations of electron binding energies are now well known \cite{Perfetti_06,Loukakos_07,Schmitt_08,Rettig_12,Papalazarou_12,Leuenberger_13}, and in a recent study on the semimetal Bi it was also shown how the underlying momentum-dependent deformation potential can be determined from such oscillations with the help of density functional theory (DFT) \cite{Papalazarou_12}. Since the electrons with the lowest binding energies determine material properties and collective phenomena, the physics will become particularly interesting if transient band shifts and renormalizations are induced near the chemical potential, which itself may then have to adjust to preserve charge neutrality. However, transient band renormalization effects in the vicinity of the chemical potential have so far only been reported for charge-density-wave systems \cite{Perfetti_06,Schmitt_08,Petersen_11,Hellmann_12}. 

\begin{figure}[t]
\includegraphics[width=\columnwidth]{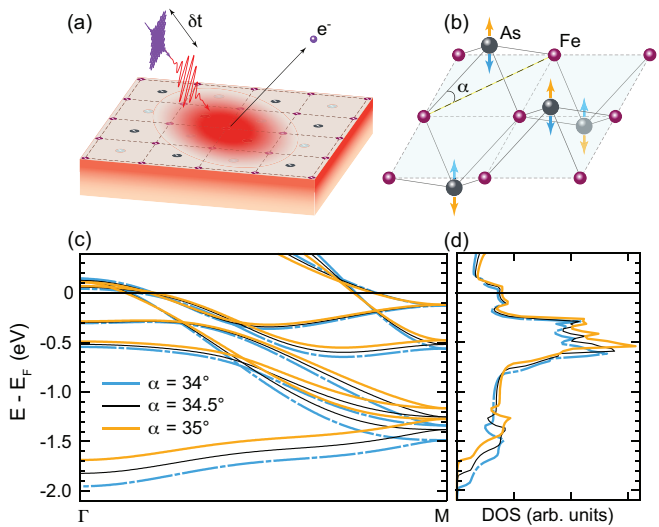}
\caption{(color online). 
(a) Schematic illustration of time-resolved pump-probe ARPES of BaFe$_2$As$_2$. The photoexcited surface region, which is out of equilibrium with respect to the bulk, is indicated.
(b) Sketch of the Fe-As layer in BaFe$_2$As$_2$ illustrating the $A_{1g}$(As) phonon mode.
(c) 10-orbital tight-binding band structure of BaFe$_2$As$_2$ at low energies along $\Gamma$--$M$ computed for various Fe-As bond angles $\alpha$ ($\alpha=34.5^\circ$ is the equilibrium value).
(d) Corresponding density of states as a function of $\alpha$.
}
\label{fig1}
\end{figure}

Iron pnictides should provide a fertile field for the study of coherent phonon-induced electronic effects near the chemical potential. Firstly, their electronic, magnetic, and superconducting properties are well known to be highly sensitive to the distance between the iron and pnictogen planes in the layered structure \cite{Johnston_10}. As an example, Figs.~\ref{fig1}(b)--\ref{fig1}(d) illustrate the predicted band structure changes for BaFe$_2$As$_2$ as a function of the Fe--As bond angle. Secondly, coherent oscillations of the relevant control parameter, the $A_{1g}$(pnictogen) phonon, can indeed be generated by femtosecond optical pulses \cite{Mansart_09,Kim_12,Avigo_13}. The most remarkable example of coherent control of iron pnictide properties has recently been provided by a time-resolved THz spectroscopy experiment on BaFe$_2$As$_2$ showing that coherent $A_{1g}$(As) phonons can drive transient spin-density-wave (SDW) ordering \cite{Kim_12}. The connection to the underlying electronic structure changes, however, is not well understood.

In this Letter, we report the direct observation of coherent phonon-induced oscillations of the effective chemical potential in photoexcited BaFe$_2$As$_2$. With amplitudes as large as $125$~meV, these oscillations markedly reflect the extreme sensitivity of the iron pnictide electronic properties near the Fermi level to the coherent breathing displacement of pnictogen atoms, and they have implications on the magnetic properties. trARPES can directly probe the varying chemical potential because the equilibration between the chemical potentials of the excited surface region and the bulk via diffusive interlayer transport is much slower than the lattice vibration. Ultrafast modulation of the chemical potential is thus established as a novel form of coherent electronic structure control.

High-quality BaFe$_2$As$_2$ single crystals with a SDW transition temperature of $138$~K were grown by a self-flux method \cite{Wang_09}. trARPES was performed using 400-nm driven high-harmonic generation \cite{Popmintchev_13} and a hemispherical electron analyzer. Sample surfaces were pumped with 30-fs near-infrared (780-nm) pulses and probed with sub-10-fs extreme-ultraviolet ($22.1$-eV and $43$-eV) pulses [Fig.~\ref{fig1}(a)]. The effective time and energy resolutions were $35$~fs and $200$~meV, respectively. Absorbed pump fluences, $F_{abs}$, ranged from $0.35$ to $2.3$~mJ/cm$^2$. Samples were measured at 100~K, if not noted otherwise.

\begin{figure}[t]
\includegraphics[width=\columnwidth]{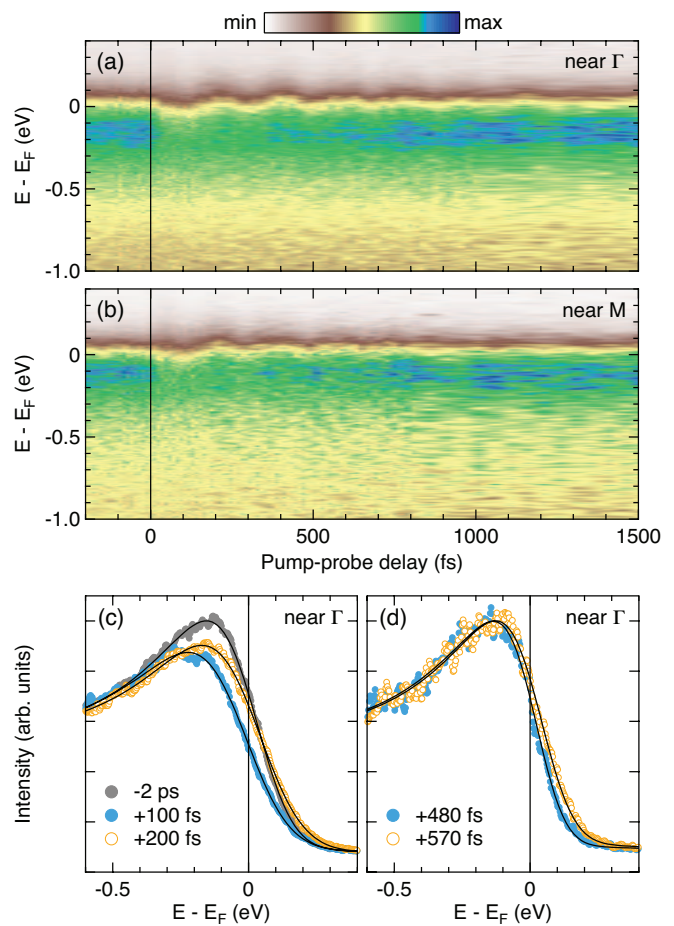}
\caption{(color online).
Time-dependent ARPES spectra of BaFe$_2$As$_2$ taken at Fermi momenta near (a) $\Gamma$ and (b) $M$. Data were recorded with $s$-polarized $22.1$-eV probe pulses at $T=100$~K and $F_{abs}=0.47$~mJ/cm$^2$. Photoemission intensity was integrated over a momentum window of $0.18$~\AA$^{-1}$. 
(c), (d) Comparison of spectra taken near $\Gamma$ at specific pump-probe delays. Solid lines represent best fits with the model described in the text.
}
\label{fig2}
\end{figure}

Figures~\ref{fig2}(a) and~\ref{fig2}(b) show the pump-induced temporal evolution of spectra taken at $F_{abs}=0.47$~mJ/cm$^2$ at momenta near $\Gamma$ and $M$ where the bands cross the Fermi level. The variation of the photoemission intensity is restricted to energies near the Fermi level. After the arrival of the pump pulse, the spectral weight in the region of maximum intensity around $-0.15$~eV is suppressed and the position of the high-energy (leading) edge starts oscillating with a sizeable amplitude of about $25$~meV, while the low-energy tail at about $-0.5$~eV stays rigid. Notably, the oscillations near $\Gamma$ and $M$ are in phase and their frequency of $(5.4\pm0.2)$~THz [period of $(185\pm6)$~fs] corresponds to the frequency of the $A_{1g}$(As) phonon mode \cite{Mansart_09,Kim_12,Avigo_13}. The spectra at selected pump-probe delays shown in Figs.~\ref{fig2}(c) and \ref{fig2}(d) indicate that the depletion of spectral weight, the modulation of the leading-edge position, and the rigidity of the low-energy tail are related to a shift and broadening of the Fermi edge as the dominant spectral change after photoexcitation.

We have performed systematic trARPES measurements at different positions in the Brillouin zone [near $X$ and near halfway $\Gamma$--$Z$ (with a probe photon energy of $43$~eV)], at different polarizations of the probe beam ($s$ versus $p$) thereby selecting hole- and electron-like bands of different symmetries, at different temperatures (100~K versus 175~K), and at different doping levels (BaFe$_2$As$_2$ versus Ba$_{0.65}$K$_{0.35}$Fe$_2$As$_2$), see below for the analysis of the results. All the results show a qualitatively and quantitatively similar temporal and spectral response indicating that a global property of the electronic structure is probed in the experiments rather than a momentum and band selective process. We argue here that this property is the effective chemical potential, $\mu$, of the photoexcited surface region of the BaFe$_2$As$_2$ sample.

\begin{figure}[t]
\includegraphics[width=0.94\columnwidth]{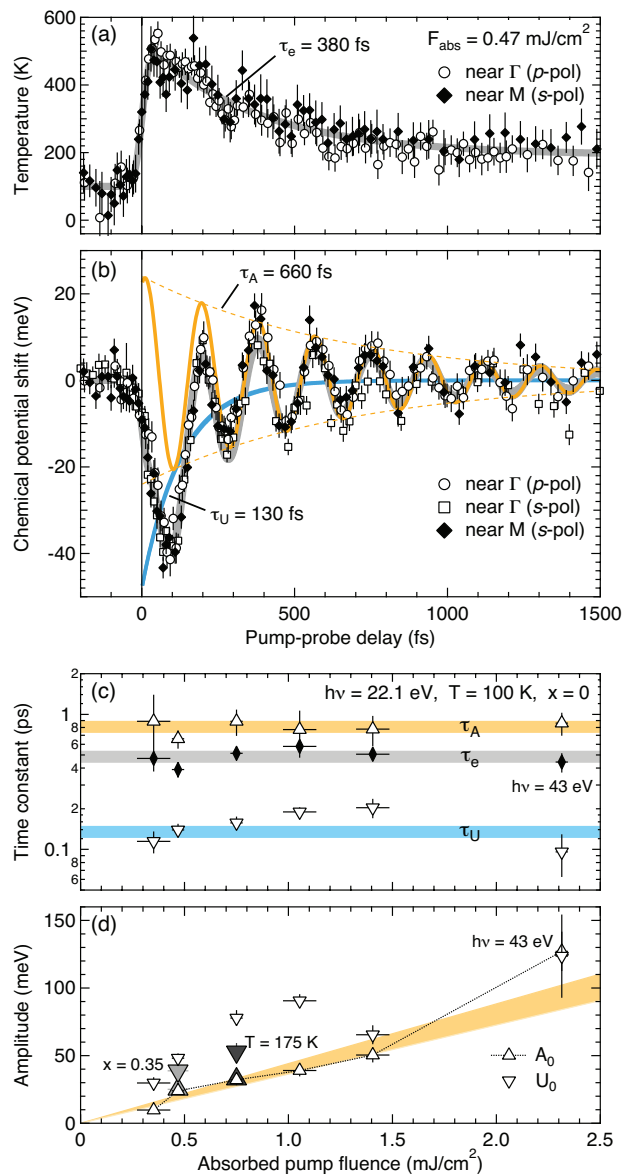}
\caption{(color online).
(a) Effective electron temperature and (b) chemical potential shift in BaFe$_2$As$_2$ as a function of pump-probe delay, extracted from trARPES data taken near $\Gamma$ and $M$ with $s$- and $p$-polarized $22.1$-eV probe pulses ($T=100$~K, $F_{abs}=0.47$~mJ/cm$^2$). Thick gray lines represent best fits with the models described in the text. In (b) oscillatory and non-oscillatory components of the temporal response are indicated.
(c) Extracted time constants and (d) amplitudes as a function of absorbed pump fluence. Lines serve as guides to the eye. In (d) filled gray symbols represent amplitudes obtained from data taken at a higher temperature and of the optimally electron-doped compound. 
}
\label{fig3}
\end{figure}

To quantify the shift and broadening of the leading edge, we have approximated the spectra with a Lorentzian plus a constant background, multiplied by a Fermi-Dirac function and convoluted with a Gaussian representing the experimental energy resolution. This implies the common assumption that electron-electron scattering is efficient on a time scale of $\leq$$100$~fs in keeping the nonequilibrium electron distribution close to an effective equilibrium distribution. As shown in Figs.~\ref{fig2}(c) and \ref{fig2}(d), the simple model fits reproduce the data very well. We emphasize that despite the limited energy resolution the fitting scheme reveals a robust modulation of the Fermi level (i.e., $\mu$): a variable $\mu$ is absolutely necessary to produce good fits, whereas a variable Lorentzian position does not significantly improve the fit quality and, when $\mu$ is fixed, cannot reproduce the experimental data. At the sensitivity of the present experiment, spectral peak shifts $\gtrsim$$20$~meV can be ruled out. Residual coherent peak shifts below this level, however, may explain the weak imprint of the oscillatory component of the temporal response on the extracted effective temperature in Fig.~\ref{fig3}(a).

Figures~\ref{fig3}(a) and \ref{fig3}(b) show the time dependencies of the effective electron temperature and the relative shift of the effective chemical potential as obtained from data taken at $F_{abs}=0.47$~mJ/cm$^2$. The effective electron temperature shoots up from $100$~K to about $500$~K and decays exponentially with a time constant, $\tau_e$, of $380$~fs. These numbers are consistent with the ones reported for a similar pump fluence in Refs.~\onlinecite{Avigo_13,Rettig_13}. The effective chemical potential, on the other hand, displays a temporal response with two components, well known from coherent phonon studies \cite{Ishioka_10}: a damped harmonic oscillation plus an exponentially relaxing shift. We model this response using the following expression for the shift of $\mu$ relative to the Fermi level of the spectrometer, $E_F$:
\begin{displaymath}
\mu(t) - E_{F} = A_0 \, \cos(\omega_A t + \phi_A) \,\mathrm{e}^{-t/\tau_A} - U_0  \, \mathrm{e}^{-t/\tau_U} 
\end{displaymath}
with the amplitudes $A_0$ and $U_0$, decay constants $\tau_A$ and $\tau_U$, and the frequency and phase $\omega_A$ and $\phi_A$, respectively. In the fitting process, $\omega_A$ is set to the frequency of the $A_{1g}$(As) mode and the finite time resolution of the experiment is taken into account by a Gaussian convolution. The best fit as well as the decomposition into the two components are included in Fig.~\ref{fig3}(b). Figures~\ref{fig3}(c) and \ref{fig3}(d) show the fit results for the time constants $\tau_A$, $\tau_U$, and $\tau_e$ and for the amplitudes $A_0$ and $U_0$ as a function of $F_{abs}$. The time constants show no systematic variation in the range of fluences applied; the distinct average values are $\langle\tau_A\rangle = (810\pm90)$~fs, $\langle\tau_U\rangle = (135\pm40)$~fs, and $\langle\tau_e\rangle = (485\pm65)$~fs. The two amplitudes, on the other hand, increase with increasing $F_{abs}$ up to a value of $125$~meV at $2.3$~mJ/cm$^2$, but they do not move in perfect correlation and $U_0$ tends to be larger than $A_0$. The extracted phase $\phi_A$ is $(-0.11\pm0.02)\pi$, independent of the fluence. This corresponds to a cosine-like oscillation as expected for a displacive excitation of coherent phonons \cite{Ishioka_10}. Note that in Fig.~\ref{fig3}(d) we have also included the amplitude values obtained from data taken at $175$~K and of optimally electron-doped Ba$_{0.65}$K$_{0.35}$Fe$_2$As$_2$.

Overall, the results of Figs.~\ref{fig2} and \ref{fig3} clearly suggest a modulation of $\mu$ that quasi-adiabatically follows the coherent oscillation of the $A_{1g}$(As) phonon mode. This, however, is surprising. On the one hand, one can see from Figs.~\ref{fig1}(c) and \ref{fig1}(d) that a small variation of the Fe-As bond angle $\alpha$ can indeed induce a significant change of the occupied band width (i.e., $\mu$). But on the other hand, BaFe$_2$As$_2$ is a metal and one would expect that $\mu$ stays put at the spectrometer reference $E_F$ and that the band positions are shifting instead, as, e.g., observed in the semimetal Bi \cite{Papalazarou_12}. To explain the effect, we need to compare the oscillation period of the coherent phonons to the time scale on which diffusive transport restores a chemical potential equilibrium between the photoexcited surface region and the bulk [Fig.~\ref{fig1}(a)]. The equilibration time is roughly given by $L^2/\pi^2 D$, where $L\approx30$~nm is a typical optical penetration depth at 780~nm \cite{Torchinsky_11} and $D=\frac{1}{3} v_{F\perp}^2 \tau_c$ is the diffusivity perpendicular to the surface ($v_{F\perp}$: Fermi velocity perpendicular to the surface, $\tau_c$: Drude relaxation time) \cite{Gamaly_13}. Around room temperature, a typical value for $\tau_c$ is $1$~fs \cite{AshcroftMermin_76}. Thus, the significantly different $v_{F\perp}$ values for BaFe$_2$As$_2$ ($\approx$$3\times10^4$~m/s \cite{Koitzsch_09,Analytis_09}) and Bi ($\approx$$10^6$~m/s \cite{AshcroftMermin_76}) translate into a three orders of magnitude difference in the equilibration time: $\approx$$300$~ps for BaFe$_2$As$_2$ versus $\approx$$270$~fs for Bi. So, while in Bi the diffusive equilibration happens fast, with a time constant comparable to the $A_{1g}$ period ($337$~fs \cite{Papalazarou_12}), any chemical potential difference between the excited surface region and the bulk in BaFe$_2$As$_2$ should indeed persist on the time scale of the coherent $A_{1g}$(As) vibration ($185$~fs). 

It is evident from the results in Fig.~\ref{fig3} that the transient modulation of $\mu$ in BaFe$_2$As$_2$ does not only depend on a coherent nuclear displacement. For an instantaneous, purely displacive excitation of coherent phonons, the amplitudes of the oscillatory and non-oscillatory components of the response, $A_0$ and $U_0$, should match, but here an additional effect shifts $\mu$ towards lower values [Fig.~\ref{fig3}(d)]. The difference is commonly attributed to electron temperature effects \cite{Ishioka_10,Papalazarou_12,Bothschafter_13}. However, in the present case the dynamics of the non-oscillatory component does not follow the dynamics of the electron temperature, as can be seen from the distinct time constants $\tau_U$ and $\tau_e$ in Fig.~\ref{fig3}(c). This specifically rules out an explanation in terms of a temperature-induced $\mu$ shift due to fine structure in the density of states near $E_F$, as put forward in a recent ARPES study \cite{Brouet_13}. Moreover, since band structure theory consistently predicts a negative slope of the density of states at the Fermi energy for BaFe$_2$As$_2$ \cite{Singh_08,Yin_11,Brouet_13}, $\mu$ would have to shift towards higher values, contrary to what is experimentally observed. Further scrutiny is required to determine the origin of the non-oscillatory component of the $\mu$ modulation. Possible contributions to this component are  the depopulation of electronic states near the Fermi level and the modification of interband interactions after photoexcitation \cite{Benfatto_11}.

Our results indicate that the easy-to-detect oscillatory component of the $\mu$ modulation can be a sensor for the coherent phonon-induced electronic structure changes near the Fermi level [Fig.~\ref{fig1}(c)] that are hidden behind the limited energy resolution. Since the data taken at a temperature and a doping level where the SDW is suppressed show unchanged amplitudes of the oscillatory component [Fig.~\ref{fig3}(d)], we conclude that a possible modulation of the SDW gap has only a minor effect on the $\mu$ oscillation. Then, conversely, using known changes to the band structure upon variation of the pnictogen height $h$ above the iron plane, we can estimate the amplitude of the coherent $A_{1g}$(As) oscillation from our results. DFT predicts a `deformation potential' $\Delta \mu / \Delta h$ for BaFe$_2$As$_2$ of about $-2$~eV/\AA\ \cite{Singh_08,Dhaka_13}. The typical renormalization factor of 3--5 for iron pnictide bands calculated within the local density approximation \cite{Yi_09,Yang_09,Werner_12} brings this value into good agreement with an experimental value of $-0.58$~eV/\AA\ derived from ARPES band shifts and measured changes of the average pnictogen height in BaFe$_2$(As$_{1-x}$P$_x$)$_2$ \cite{Yoshida_11,Ye_12,Rotter_10}. The amplitude of the $\mu$ oscillation measured at $F_{abs}=0.47$~mJ/cm$^2$ is $24$~meV. We thus obtain an amplitude of the coherent As displacement of about $4$~pm, corresponding to a modulation of the bond angle $\alpha$ of $0.8^\circ$. These values are by a factor of 5 larger than the ones estimated in a recent time-resolved THz spectroscopy study \cite{Kim_12}, but the displacement compares well with the amplitude of coherent $A_{1g}$ oscillations in Bi derived from trARPES data \cite{Papalazarou_12}, all studies performed at similar pump fluences.

Notably, the observed initial decrease of the effective chemical potential [Fig.~\ref{fig3}(b)] corresponds to an initial increase of the As height which goes along with an increase of the density of states at the Fermi level [Fig.~\ref{fig1}(d)]. The initial direction of the coherent As motion should therefore correlate with an enhanced tendency towards SDW order. From the results of DFT calculations of BaFe$_2$As$_2$ \cite{Rotter_10} as well as from the results of magnetic and structural measurements of BaFe$_2$(As$_{1-x}$P$_x$)$_2$ \cite{Iye_12,Rotter_10}, we can derive a rate of change of the Fe magnetic moment with the average pnictogen height of $\Delta m / \Delta h \approx 10$~$\mu_B$/\AA. A coherent As breathing vibration with an amplitude of $4$~pm can thus induce a transient change of the magnetic moment as large as $\pm0.4$~$\mu_B$, corresponding to a relative change of $\pm43$\% \cite{Wilson_09}. In consequence, in the first half-cycle the amplitude of magnetic fluctuations may be significantly enhanced. However, it remains to be understood how the coherent phonons drive the fluctuations into coherence to develop transient SDW order above the transition temperature \cite{Kim_12}.

In conclusion, the chemical potential of photoexcited BaFe$_2$As$_2$ is shown to oscillate at the frequency of the $A_{1g}$(As) phonon mode. This observation substantiates electronic structure control of the iron pnictides by coherent phonons, with further implications on the magnetic properties of these materials. Persistent chemical potential modifications could be generic to layered materials with weak interlayer coupling when driven out of thermal and diffusive equilibrium.

This work was supported by the German Science Foundation (DFG) through project BA~2177/9-1 and the priority program SPP~1458. L.X.Y.\ and R.C.\ thank the Alexander von Humboldt foundation for support. B.K.\ and I.E.\ acknowledge support by the Mercator Research Center Ruhr (MERCUR).

\end{document}